\documentclass{llncs}
\usepackage[T1]{fontenc}
\usepackage[colorinlistoftodos, obeyDraft]{todonotes} 
\usepackage{graphicx}
\usepackage{hyperref}
\usepackage{natbib}
\usepackage{subcaption}
\usepackage{amsmath}
\usepackage{float}

\usepackage{multirow}
\usepackage{makecell}

\usepackage{siunitx}                        
\usepackage{threeparttable}  
\usepackage{color}

\begin{document}
\title{How Exclusive are Ethereum Transactions? \\ Evidence from non-winning blocks\thanks{We are grateful to Agnostic Relay for providing the data, especially to Ombre for patiently explaining how relays work.  We are also grateful to  Agostino Capponi, Hanna Halaburda, Fahad Saleh, the participants to the workshops ``(Back to) The Future(s) of Money II'' and the 2025 ``CBER CtCe'' conference for their comments and suggestions. }}
\titlerunning{How Exclusive?}
%
\author{Vabuk Pahari\inst{1}  \and Andrea Canidio\inst{2} 
}
\authorrunning{Pahari and Canidio}
%
\institute{$^2$ CoW Protocol, 
\email{andrea@cow.fi}\\
$^{1}$ Max-Planck-Institute for Software Systems, 
\email{vpahari@mpi-sws.org}
\\
~
\\
\today.
}

\maketitle              
\begin{abstract}
We analyze 15,097 blocks proposed for inclusion in Ethereum’s blockchain over an eight-minute window on December 3, 2024, during which 38 blocks were added to the chain. We classify transactions as \textit{exclusive}---appearing only in blocks from a single builder---or \textit{private} --- absent from the public mempool but included in blocks from multiple builders. We find that, depending on the methodology, exclusive transactions account for between 77.2\% and 84\% of the total fees paid by transactions in winning blocks. Moreover, we show that exclusivity cannot be fully attributed to persistent relationships between senders and builders: only between 7\% and 8.4\% of all on-chain exclusive transaction value originates from senders who route exclusively to one builder. Finally, we observe that transaction exclusivity is dynamic. Some transactions are exclusive at the start of a bidding cycle but later appear in blocks from multiple builders. Other transactions remain exclusive to a losing builder for two or three cycles before appearing in the public mempool. These transactions are therefore delayed and then exposed to potential attacks.
\end{abstract}

\keywords{Ethereum, Proposer-Builder Separation (PBS), Order Flow.}
%

%
\section{Introduction}

Ethereum separates the roles of the block \textit{proposer} and the block \textit{builder}. The proposer is randomly selected from users who have staked at least 32 ETH, and is responsible for adding the next block to the chain. Builders, by contrast, are professional entities that gather transactions from both the public mempool and private channels, and use sophisticated algorithms to decide which transactions to include in a block and in what order. They then compete in an auction by submitting multiple candidate blocks, each with a payment to the proposer in case the block is added to the blockchain. This auction is highly competitive, with thousands of candidate blocks for each block selected for inclusion in Ethereum's blockchain.

This system, known as \textit{Proposer-Builder Separation} (PBS), has proven effective at selecting the most valuable transactions for inclusion on chain. Its main drawback, however, is the resulting concentration in the builders’ market: at present, the top builder (Titan) produces about 50\% of all blocks, while the top three account for over 90\%.\footnote{For up-to-date statistics, see \url{https://www.relayscan.io/}. Buildernet is a joint project by Flashbots (which operated its own independent builder until mid-2025) and Beaverbuild (which both participates in Buildernet and runs its own builder).} This concentration grants a few entities disproportionate influence over which transactions are included, undermining Ethereum’s foundational principles of decentralization and permissionlessness.  

The prevailing explanation for the concentration observed under PBS centers on \textit{exclusive} transactions, i.e., those available to a single builder only. Transactions shared among multiple builders should not affect the equilibrium outcome of the PBS auction, since all builders should raise their proposed payments to the proposer by an amount equal to these transactions' fees, which ultimately accrue to the winning builder. As a result, the auction outcome is primarily determined by exclusive transactions. Furthermore, the more frequently a builder wins, the more attractive it becomes as a destination for non-shared order flow, further increasing its likelihood of winning future auctions and leading to concentration in the builders’ market (see \citealp{Quintus}).

Despite their central role in the PBS mechanism, quantifying the importance of exclusive transactions has proved difficult, since on-chain data alone cannot reveal whether a transaction was privately shared among multiple builders or only with the winner. Our paper takes a first step toward addressing this challenge using a novel dataset comprising 15{,}097 blocks \textit{proposed} for inclusion in Ethereum over an eight-minute window on December 3, 2024, during which 39 new blocks were added.\footnote{The dataset includes all blocks (winning and non-winning) submitted via the ``Agnostic Relay’’ during the study period. They are therefore not a random sample of all proposed blocks (see Section~\ref{sec: dataset description} for details and limitations). To our knowledge, together with the companion paper~\cite{canidio2025becoming}, this is the first study to analyze the content of non-winning Ethereum blocks.} The dataset contains 10{,}793 unique transactions (identified by hash) and 2{,}380{,}014 transaction–block pairings. By comparing the content of proposed blocks with the public mempool (sourced from \url{https://mempool.guru/}), we classify each transaction as:
\begin{itemize}
    \item \textbf{Public}: appeared in the public mempool;
    \item \textbf{Private}: appeared in blocks from multiple builders but not in the public mempool;
    \item \textbf{Exclusive}: appeared only in blocks from a single builder.
\end{itemize}
We apply this classification both globally and within shorter time intervals, allowing us to track how transaction statuses evolve. For example, a transaction may initially be exclusive but later become public or private.

\paragraph{Summary of results.}
Of the 5,576 transactions in our dataset that were ultimately included on-chain, 4,124 were public, 810 private, and 642 exclusive. Exclusive transactions, though fewer in number, are by far the most valuable: they account for 84\% of all fees paid by transactions included in winning blocks. This result confirms their central role in block building. We also find that almost all (i.e., 91\%) transactions that are never included in a winning block are exclusive.

Because we identify transactions by their hash, it is possible that two exclusive transactions available to different builders perform identical or nearly identical operations, meaning they may functionally be the same transaction and thus are not truly exclusive. To examine this possibility, we study cases in which the same sender submits exclusive transactions to multiple builders within a single auction cycle. We then use their transaction logs to identify cases in which the same sender sends exclusive transaction that are:
\begin{itemize}
    \item functionally identical to each other, executing the same operation and paying the same fee;
    \item identical in execution but different in the attached fee; and
    \item similar but not identical, performing swaps in the same DEX pools but with different amounts.
\end{itemize}
While this log-based analysis refines our classification of exclusivity, the number of affected transactions is small and does not alter our main conclusion: even if all such transactions were reclassified from ``exclusive'' to ``private,'' exclusive transactions would still account 77.2\% of total fees paid by transactions included in winning blocks.

To investigate the origin of exclusive transactions, we then analyze the relationship between transaction senders and builders. Counting transactions alone, we find that the vast majority of exclusive transactions can be attributed to senders that interact exclusively with a single builder. However, when we shift focus from transaction counts to transaction \textit{value} and consider only those that are ultimately included on-chain, the picture changes. Under our preferred operational definition of a ``transaction sender,'' only about 7\% of the total value of exclusive transactions can be attributed to exclusive relationships between searchers and builders. Such relationships therefore appear to play a limited role in explaining the overall value of exclusive transactions.

Finally, we study whether transactions change status over time, starting with changes \textit{across} bidding cycles. Among transactions that first appear as exclusive, 162 remain exclusive over multiple cycles, 174 transition to private in a subsequent cycle, and 12 later appear in the public mempool. These 12 transactions appear to be user-generated, as they involve swaps on DEX pools via well-known routers such as 0x and 1inch. Their initial exclusivity is puzzling, as ordinary users typically gain no benefit from sending transactions  to a single builder instead of submitting them to a privately to multiple builders or to the public mempool. Even more surprisingly, these transactions eventually appear in the public mempool, indicating that they were initially delayed and later exposed to potential attacks (one was in fact the victim of a sandwich attack). In four cases, the same builder who initially had exclusive access ultimately included the transaction on-chain. We speculate that a private mempool operator may have shared these transactions exclusively with one builder, reverting to the public mempool when that builder failed to win.\footnote{Private mempools are services that allow users to send transactions directly to one or more builders without broadcasting them to the public mempool. During the study period, there were five major private mempool operators. Four of them --- MEV Blocker, Flashbots Protect, Merkel, and Blink --- state in their documentation that they share transactions privately with multiple builders (on the differences between those operators, see \citealp{janicot2025privatemevprotectionrpcs}). Instead, the fifth operator, Metamask Smart Transaction, does not disclose its sharing policy.} 

We also observe transactions changing status \textit{within} a single bidding cycle. Specifically, we identify 19 cases in which transactions are initially exclusive to one builder at the start of the auction cycle but later, within the same cycle, appear in blocks proposed by other builders. Most of these transactions were submitted via Flashbots Protect: they were initially exclusive to the Flashbots builder and only later propagated more broadly. Our interpretation is that Flashbots builder temporarily retains transactions submitted through Flashbots Protect as exclusive, releasing them to other builders once it becomes apparent that it is unlikely to win the block.

\paragraph{Relevant literature}

To our knowledge, \cite{yang2025decentralization} is the only other paper that examines the content of non-winning blocks. They study builder competition by comparing bids with realized block revenues (i.e., total payments received by builders), but do not analyze individual transactions. The authors also propose a heuristic to infer \emph{exclusive} transactions from bid-level data. In their approach, exclusive transactions are defined as private transactions included in a winning block from an originator who is \emph{pivotal}—that is, if the transactions from that originator were removed, the block’s value would fall below the second-highest builder’s bid (multiplied by a calibration threshold). This method effectively identifies some sources of exclusive order flow but has two main limitations. First, a set of transactions may appear pivotal yet be present also in the second-most valuable block. Second, a truly exclusive transaction may fail to appear pivotal if the runner-up block contains a different exclusive transaction of comparable value.

Other studies attempting to identify ``private'' and ``exclusive'' transactions have been limited by the difficulty of disentangling them using on-chain data alone. For example, \cite{thiery2023empirical} labels as \emph{exclusive} any transaction included on chain without first appearing in the public mempool---what we would classify as either private or exclusive---and reports that such transactions account for roughly 80\% of the total value paid to builders, consistent with our findings (a similar estimate is reported by \citealp{yang2025decentralization}). \cite{frontier} studies flows from identified searchers (bots) and documents a positive relationship between builder market share and flow from searchers, without distinguishing between private and exclusive transactions. Finally, \cite{oz2024wins} shows that builders' profitability correlates with access to exclusive transactions. However, their definition of ``exclusive'' includes all transactions not seen in the public mempool and not identified shared via MEV Blocker or Flashbots protect, two private mempool services. In our dataset, these two sources account for 320 of the 810 transactions that we classify as private.

A growing theoretical literature examines builder competition and the role of exclusive transactions in shaping the outcome of the PBS auction. In particular, \cite{gupta2023centralizingeffectsprivateorder} and \cite{capponi2024proposer} formalize how access to exclusive transactions can lead to concentration in the builder market. \cite{bahrani2024centralization} introduce heterogeneity among builders and derive conditions under which both builder and proposer markets become concentrated. \cite{wu2024strategic,wu2024compete} analyze builders’ bidding strategies, combining theoretical modeling with simulation-based validation. \cite{lehar2023battle} and \cite{CAPPONI2025104132} study the tradeoff between submitting transactions via the public mempool versus private channels, and characterize the resulting equilibrium sorting of transactions; \cite{CAPPONI2025104132} also provides empirical support for its model using data from Flashbots Protect. Similarly, \cite{azar2024information} develop a theory in which the choice between public and exclusive submission depends on the exogenous value of private information; when this value is sufficiently high, the model predicts concentration in the builder market. They test this theory empirically by classifying transactions as ``public'' or ``private,'' under the assumption that private transactions are available exclusively to a single builder.

Finally, in our companion paper \cite{canidio2025becoming}, we use the same dataset to address two related questions: whether block selection matters for users, and competition between arbitrageurs.
We shows that many user transactions are \emph{delayed}: although proposed during a bidding cycle, they are not included in the winning block. Furthermore, we identify two arbitrage bots trading between decentralized (DEX) and centralized (CEX) exchanges. Analyzing the bidding dynamics, they estimate that the implied CEX prices at which they trade USDC/WETH and USDT/WETH is 2.8 basis points \textit{better} than the contemporaneous Binance price.

\section{Background on PBS}

As noted above, in Ethereum, the proposer almost always delegates block construction to builders. Builders collect both user transactions and those submitted by \textit{searchers}---automated bots that generate transactions in response to market conditions (e.g., price fluctuations) or to other transactions pending in the public mempool. Searchers play a central role in decentralized finance: they engage in arbitrage across DEXs and CEXs, liquidate under-collateralized loans, or exploit temporary price discrepancies between DEXs. Because their strategies often depend on speed and precise ordering, searchers typically route transactions to builders via private channels. 

Users can submit transactions in different ways. The default is to broadcast them to the public mempool, where all builders can observe them. This, however, exposes users to attacks.\footnote{For example, if a user swaps assets on an Automated Market Maker (AMM), a malicious searcher can mount a ``sandwich'' attack: front-running the swap with the same trade and then back-running it with the opposite trade. The attacker buys low and sells high, forcing the victim to execute at worse prices.} To reduce this risk, many users rely on private mempools, publicly available services which distribute transactions privately to a subset of builders.

The relation between builders and proposers is facilitated by \textit{relays}. Their primary role is to prevent proposers from altering the submitted blocks.\footnote{Technically, this is achieved by requiring proposers to sign an empty block, which is then completed by the relay. For more details, see \url{https://docs.flashbots.net/flashbots-mev-boost/relay}.} Once a builder creates a block, it submits it to a relay along with a proposed payment to the proposer. The relay then forwards the highest-paying block and its corresponding payment to the proposer upon request. This process resembles an ascending price auction because the relay continuously broadcasts the value of the highest-paying bid---though this broadcast is observed with a delay due to network latency. An auction cycle lasts approximately 12 seconds, after which the winning block is selected, and a new auction begins. During each auction cycle, builders continuously resubmit blocks, modifying the transactions included and bid amounts. Latency plays a crucial role, as proposers are geographically dispersed, and arbitrage transactions often depend on rapid price movements in traditional ``off-chain'' financial markets. To account for geographic dispersion and fast-moving arbitrage opportunities, multiple relays operate worldwide, with builders submitting to several at once. Our dataset, for instance, comes from a relay called Agnostic Relay.

\section{Dataset}\label{sec: dataset description}

Our primary dataset consists of \textit{all} blocks (winning and non-winning) submitted via Agnostic relay between Ethereum blocks 21,322,622 and 21,322,660 (slots 10,534,387--10,534,425), spanning 2:37:35--2:45:25 PM UTC on December 3, 2024.\footnote{Founded in 2022, Agnostic relay held about 20\% of the relay market in 2023, declining to around 5\% during our study period.} We complement this dataset with several public sources:
\begin{itemize}
    \item All blocks added to the Ethereum blockchain during our study period, including those not submitted via Agnostic relay. We also track ten additional winning blocks beyond the study period to see whether transactions in our dataset were eventually included on-chain.  
    \item Hashes of blocks submitted to all major relays: Flashbots, BloXroute, Manifold, Eden, Ultra Sound, SecureRpc, and Aestus. Because any difference in content or bid produces a unique hash, this lets us check whether blocks in our dataset were also sent elsewhere.
     \item Data on transactions submitted via the public mempool, obtained from \url{https://mempool.guru/}.
    \item Data on transactions submitted via two private mempools, MEV Blocker and Flashbots Protect, obtained from \url{https://Dune.com}.
\end{itemize}
Finally, we simulate selected transactions in our dataset to test whether transactions with different hashes nonetheless execute the same or similar operations. These simulations are run on an Ethereum Archive Node using the Erigon client, forking at the relevant block heights.

An important detail of PBS for our analysis is that a relay does not run an auction for a given slot if the assigned proposer is not registered with it. This occurs in our data: of the 39 slots we study, 11 show no bidding activity because the proposers were not connected to the Agnostic relay.\footnote{These are the slots leading to blocks 21,322,624; 21,322,625; 21,322,627; 21,322,629; 21,322,633; 21,322,634; 21,322,636; 21,322,644; 21,322,646; 21,322,647; and 21,322,651.} By comparing the hash of blocks in our primary dataset with those submitted to the other major relays, we find that our 15,097 blocks represent 28.3\% of all blocks submitted to major relays during the 28 auction cycles for which we have data.

Also, each block in our dataset has two timestamps: \textit{received at} (i.e., received from the builder) and \textit{made available at} (i.e., made available to the proposer). The gap between them reflects the time needed for the relay to simulate the block and verify its validity.\footnote{Some blocks are treated ``optimistically''---they are made available immediately, with simulation performed in the background. If the simulation later fails, the builder’s next submission is not treated optimistically. Such blocks account for only 3.8\% of our sample.} This simulation delay is non-trivial: the median is 0.76 seconds, with the 75th percentile at 1.5 seconds. In what follows, we use the \textit{received at} timestamp, as it better captures the events we study---such as when a builder submits a block or when a transaction first appears.

As noted earlier, our primary dataset covers 15,097 blocks submitted across 28 bidding cycles. Across those cycles, the number of blocks per slot ranges from 220 to 951, with a mean of 539 and a median of 509.  We can attribute 14,043 of these blocks to 23 known builders. The most active are Titan Builder (7,024 blocks, 46.5\%), Rsync Builder (2,259, 15.6\%), Flashbots (1,936, 12.8\%), and Beaverbuild (1,418, 9.4\%). Most blocks in our dataset---12,895 in total---were also submitted to at least one other relay.  However, when comparing across \textit{all} relays, we find an imbalance. Titan Builder submitted 13,679 blocks overall (25.6\%), Rsync Builder 7,936 (14.9\%), Flashbots 2,626 (4.9\%), and Beaverbuild 21,447 (40.2\%). Thus, Beaverbuild contributes far fewer blocks to Agnostic Relay than to other relays. We recognize this imbalance as a limitation of our dataset.

Just as searchers and users pay builders to include their transactions, builders pay proposers through a \textit{fee reception} transaction appended at the end of each block. In our dataset of 15,097 blocks, 13,848 (91.7\%) contain such a transaction, corresponding to 12,205 distinct fee reception transactions (different blocks may include the same fee reception transaction).  The remaining blocks---those without a fee reception transaction---are low-value submissions in which the builder sets the \textit{fee recipient} variable directly to the proposer’s address. In these cases, the bid equals the sum of payments sent to that address.  In what follows, we use the term ``transactions'' to mean all transactions \textit{excluding} fee reception transactions. This yields 10,793 unique transactions (by hash), of which 5,576 were eventually included in a winning block, either in the cycle where they first appeared or later.

\section{Results}

\begin{figure*}[t!]
    \centering
    \hspace{-40mm}
    \begin{subfigure}[t]{0.55\textwidth}
        \centering
        \includegraphics[scale=0.3]{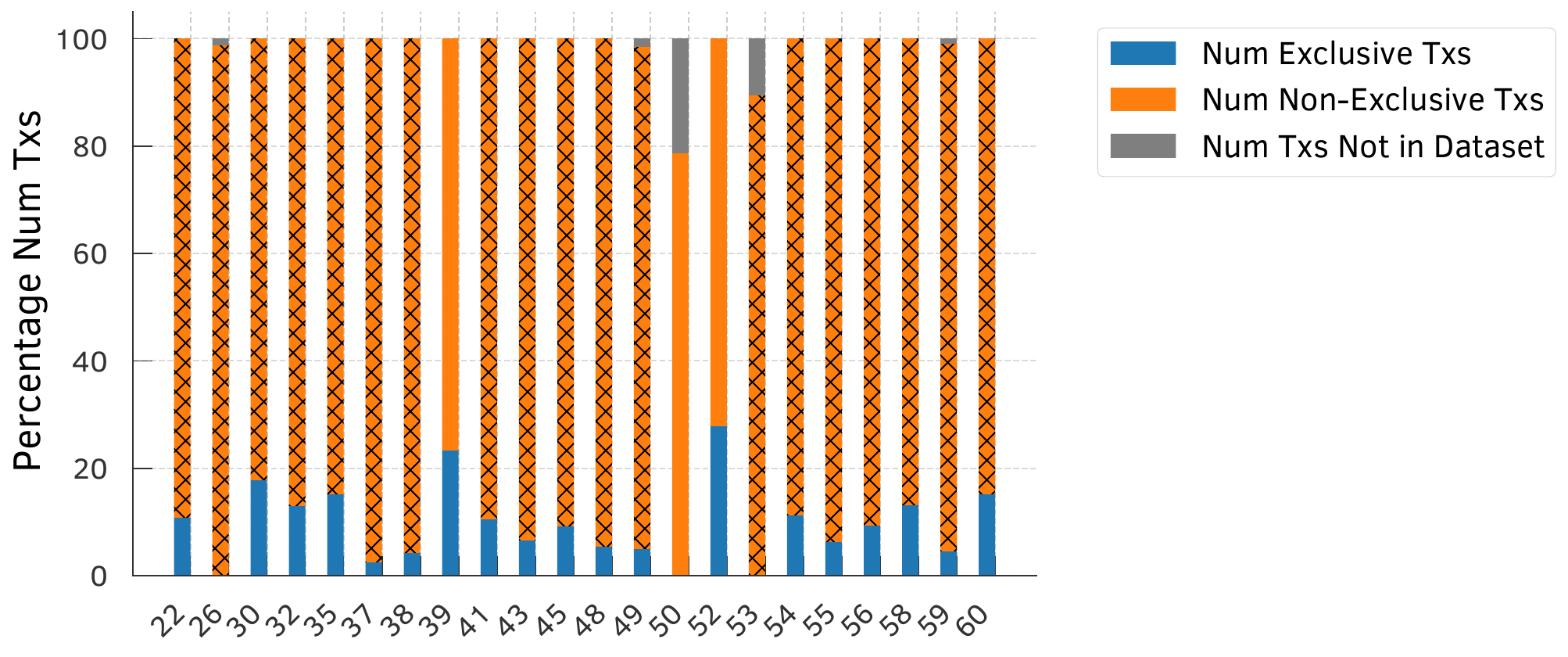}
        \caption{Percentage of Transaction}
        \label{fig:numtxs-blocks}
    \end{subfigure}%
    \begin{subfigure}[t]{0.3\textwidth}
        \centering
        \includegraphics[scale=0.3]{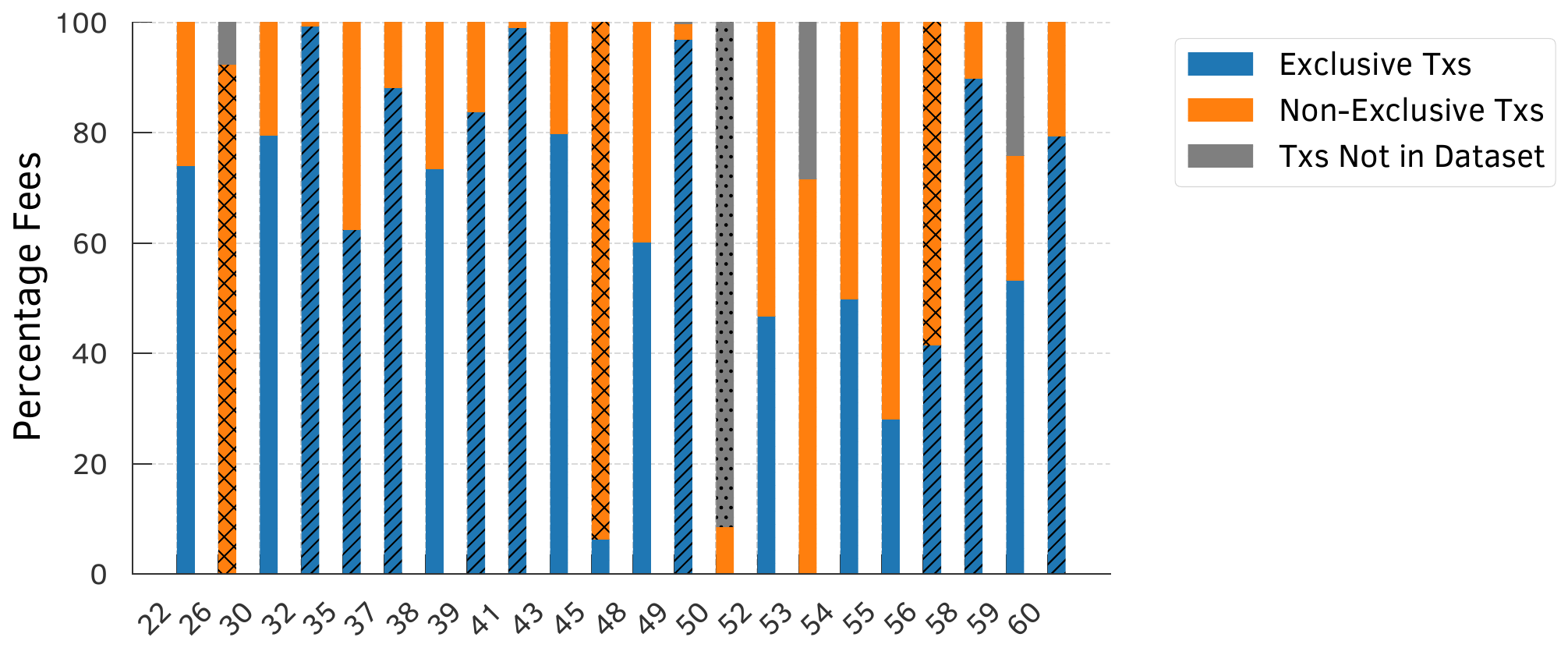}
        \caption{Percentage of Fees}
        \label{fig:fees-blocks}
    \end{subfigure}%
    \caption{Exclusive, non-exclusive (public and private), and non-identified transactions for each winning block. For each block, we report both the share of transactions and the share of total fees attributable to each category. Blocks are numbered 21{,}322{,}622 through 21{,}322{,}660.
    }
    \label{fig:fee-paid-exclusive-txs}
\end{figure*}

We label the transactions in our dataset using MempoolGuru as either \textit{public} (it appeared in the public mempool) or \textit{non-public} (it appeared in a block but not in the mempool). We further split non-public transactions into \textit{private} (present in blocks from multiple builders) and \textit{exclusive} (present only in blocks from a single builder). Note that transaction that appear in winning blocks during the study period but do not appear in our dataset (because included only in blocks not submitted to Agnostic relay) cannot be categorized.

Focusing on the 5,576 transactions in our dataset that were ultimately included in a winning block, MempoolGuru classifies 4,124 as public and 1,452 as non-public. Of the non-public transactions, 266 were submitted via MEV Blocker, 40 via Flashbots Protect, and 14 through both---indicating that these are private but not exclusive. This leaves 1,132 non-public transactions to classify further. Using our dataset, we find that 642 of them (45\% of all non-public transactions in winning blocks) are exclusive: 533 appear only in Titan Builder’s blocks, 107 only in Rsync Builder’s blocks, and 2 only in builder0x69’s blocks.  

In value terms, the dominance of exclusives is stark. Of the 14.67 ETH in fees paid by all transactions eventually included in winning blocks, 12.27 ETH (84\%) came from exclusive transactions, 1.66 ETH (11\%) from private ones, and 0.74 ETH (5\%) from public ones. Exclusive transactions are thus by far the builders’ primary source of revenue, reinforcing the idea that they drive the observed dynamics in the builders' market.


With respect to the 5,217 transactions never included in a winning block,  we find 4,769 exclusive transactions and 448 private transactions. Hence, the vast majority of discarded transactions are exclusive.

As noted, a limitation of our analysis is that we observe only a small share of Beaverbuild’s blocks. This creates two concerns. First, some transactions we classify as ``exclusive'' may in fact also appear in Beaverbuild blocks outside our dataset. For this to occur, a transaction would need to be shared with Beaverbuild and only one other builder in our dataset. However, sharing transactions with exactly two builders is quite rare in our dataset: among the 6,669 non-public transactions in our dataset, we find only 435 that were shared with exactly two builders. Hence, the number of transactions categorized as "exclusive" while in reality they are ``private'' is probably small. Second, and more importantly, transactions exclusive to Beaverbuild are missing from our analysis whenever Beaverbuild did not submit the corresponding block to Agnostic, even if those transactions were ultimately included on-chain. As a result, the overall share of exclusive transactions during our study window may be larger than what we measure.

To further support our hypothesis, Figure~\ref{fig:fee-paid-exclusive-txs} reports, for each winning block in our study period, the share of transactions that we classify as exclusive, as non-exclusive, or as not identifiable (because they do not appear in our dataset). Consistent with the discussion above, the only winning blocks with no exclusive transactions are the three built by Beaverbuild (21322626, 21322650, and 21322653). By contrast, the remaining 19 winning blocks---all produced by Titan or Rsync---almost never contain transactions that we cannot identify: 17 of them contain none, one block contains a single non-identified transaction of negligible value, and only one block (21322659 by Rsync) contains both economically meaningful exclusive transactions and non-identified transactions. Taken together, these patterns suggest a sharp separation across builders: Beaverbuild’s winning blocks contain non-identified transactions but no exclusive ones, whereas blocks built by the other major builders contain exclusive transactions but (almost) no non-identified ones. This supports our hypothesis that most transactions we cannot identify are exclusive to Beaverbuild. Finally, if we reclassify all transactions appearing in winning blocks but missing from our dataset as exclusive, the share of winning-block value attributable to exclusive transactions rises to 87\%.

\subsection{Logs-based classification}

The above method treats transactions with different hash as different transactions. However, a transaction may have a different hash from another while performing the same operations or similar ones. As a result, our method may overestimate the share of exclusive transactions. To address this concern, in this section we refine our approach by using transaction logs, that is, the sequence of operations and corresponding outputs generated by simulating the blocks that included those transactions.

The method above treats transactions with different hashes as distinct. However, two transactions may have different hashes while performing the same or very similar operations. As a result, our method may overestimate the share of exclusive transactions. To address this concern, we refine our approach in this section by comparing transaction logs, that is, the sequence of operations and corresponding outputs generated by simulating the blocks that included the transactions.

We identify exclusive transactions originating from the same sender but submitted to different builders within the same auction cycle. To do so, we must first define what constitutes a "sender." A natural option is the transaction’s \texttt{from} address. However, many searchers use multiple \texttt{from} addresses while embedding their logic in smart contracts that automatically generate transactions. For this reason, the \texttt{to} address—that is, the address of the smart contract called by the transaction—is a more reliable identifier of the sender, particularly in the case of searchers. Since multiple \texttt{from} addresses can call the same contract, grouping transactions by \texttt{to} address increases the likelihood of detecting functionally similar transactions that differ only in hash. In this sense, using \texttt{to} addresses offers a stricter test of our hash-based classification of transactions as exclusive, private, or public. We therefore adopt the \texttt{to} address as the definition of sender in our analysis.

 We we simulate these transactions to extract their logs, we identify 665 such transactions producing 279 unique logs. Nearly all of them (630 transactions) are swaps in DEX pools, primarily on Uniswap V2, Uniswap V3, Sushiswap, Pancakeswap, or Curve. We categorize these transactions in two groups:  
\begin{itemize}
    \item \textbf{Partially exclusive:} An exclusive transaction is deemed partially exclusive if another exclusive transaction from the same sender in the same auction cycle produces an identical log. To measure its ``degree of exclusivity,'' we compare all such transactions (same sender, same cycle, same log) and compute the difference between the highest and second-highest fee \textit{paid to different builders}. Note that the degree of exclusivity can be zero if multiple exclusive transactions share the same log and fee.\footnote{Two transactions may generate different hashes despite being functionally identical. In our dataset, we find cases of transactions with identical logs and fees but differing in gas limit (55 cases),  max priority fee (11 cases),  nonce or sender (29 cases), and transaction value (32 cases, ``transaction value'' is an additional ETH transfer that a transaction may perform).}  
    \item \textbf{Similar:} Two exclusive transactions are classified as similar if they come from the same sender in the same auction cycle, swap in the same DEX pools, but differ in the amounts swapped, resulting in distinct logs.  
\end{itemize}

We find 171 partially exclusive transactions (59 unique logs) with a degree of exclusivity equal to zero---that is, the sender submitted the same transaction (same log, same fee) to at least two builders. We found another  494 partially exclusive transactions (220 unique logs) with different fees across builders. Among these, the average degree of exclusivity is 0.004178 ETH (or 23\% of the value of the second-highest fee). We also identify 1,179 ``similar'' exclusive transactions grouped into 162 log sets (861 unique logs), where transactions differ only in the swap amount. Comparing the fees attached to these transactions, we find that the average gap between the highest payment to one builder and the second-highest payment to another builder is almost 70\% of the second-highest payment. This suggests that the differences between these transactions are economically meaningful.

%


The takeaway is that not all transactions that appear ``exclusive'' at the hash level remain exclusive once we examine their logs: some are duplicates submitted to multiple builders, while others are variations of the same strategy. Quantitatively, however, these cases are limited. Also, if we change the classification of partially exclusive and similar transactions to ``private'', we find that 77.2\%  of fees paid by transactions that are included on-chain are from exclusive transactions (vs 84\% with our initial classification). The main result, therefore, stands: exclusive transactions remain the key driver of builders' profitability. 


\subsection{Relationship between searchers and builders}

To better interpret these results, we study whether exclusive transactions are the outcome of exclusive relationships between senders and builders, or, instead, the same sender submits different exclusive transactions to different builders. 


\begin{table}[htb]
\centering
\begin{tabular}{||c||c|c|c|c||c|c|c|c||} 
 \hline
 \multirow{2}{*}{\# of builders} 
   & \multicolumn{4}{c||}{Transactions} 
   & \multicolumn{4}{c||}{Fees (ETH)} \\ 
 \cline{2-9}
   & From & To & From $\geq$5 & To $\geq$5 
   & From & To & From $\geq$5 & To $\geq$5 \\ 
 \hline\hline
 1 & \makecell{3,338 \\ (63.85\%)} & \makecell{3,053 \\ (58.4\%)} & \makecell{2,757 \\ (60.58\%)} & \makecell{2,764 \\ (56.35\%)} 
  & \makecell{80.6 \\ (59.2\%)} & \makecell{66.7 \\ (49.0\%)} & \makecell{60.55 \\ (53.1\%)} & \makecell{58.8 \\ (51.6\%)} \\ 
 \hline
 2 & \makecell{1,356 \\ (25.9\%)} & \makecell{1,455 \\ (27.83\%)} & \makecell{1,264 \\ (27.8\%)} & \makecell{1,421 \\ (28.97\%)} 
   & \makecell{35.1 \\ (25.8\%)} & \makecell{39.6 \\ (29.1\%)} & \makecell{32.9 \\ (28.9\%)} & \makecell{38.1 \\ (33.46\%)} \\
 \hline
 3 & \makecell{281 \\ (5.3\%)} & \makecell{87 \\ (1.66\%)} & \makecell{277 \\ (6.09\%)} & \makecell{87 \\ (1.77\%)} 
   & \makecell{15.8 \\ (11.6\%)} & \makecell{20.3 \\ (14.9\%)} & \makecell{15.8 \\ (13.86\%)} & \makecell{20.3 \\ (17.8\%)} \\
 \hline
 4 & \makecell{87 \\ (1.6\%)} & \makecell{92 \\ (1.76\%)} & \makecell{87 \\ (1.91\%)} & \makecell{92 \\ (1.88\%)} 
   & \makecell{0.85 \\ (0.63\%)} & \makecell{1.02 \\ (0.75\%)} & \makecell{0.85 \\ (0.75\%)} & \makecell{1.02 \\ (0.89\%)} \\
 \hline
\end{tabular}
\vspace{.1cm}
\caption{Exclusive transactions by sender, grouped by how many distinct builders each sender interacts with. 
Columns distinguish senders defined by \texttt{from} vs. \texttt{to} addresses, with and without a minimum threshold ($\geq 5$ transactions). The left panel reports transaction counts with the share of total transactions underneath. The right panel reports total fees in ETH with the share of total fees underneath.}
\label{table:exclusive_relationship_and_fees}
\end{table}

\begin{table}[htb]
\centering
\begin{tabular}{||c||c|c|c|c||c|c|c|c||} 
 \hline
 \multirow{2}{*}{\# of builders} 
   & \multicolumn{4}{c||}{Transactions} 
   & \multicolumn{4}{c||}{Fees (ETH)} \\ 
 \cline{2-9}
   & From & To & From $\geq$5 & To $\geq$5 
   & From & To & From $\geq$5 & To $\geq$5 \\ 
 \hline\hline
 1 & \makecell{366 \\ (74.7\%)} & \makecell{331 \\ (67.6\%)} & \makecell{46 \\ (30.1\%)} & \makecell{162 \\ (51.3\%)} 
   & \makecell{5.96 \\ (48.6\%)} & \makecell{0.94 \\ (7.7\%)} & \makecell{2.03 \\ (24.64\%)} & \makecell{0.87 \\ (7.12\%)} \\ 
 \hline
 2 & \makecell{97 \\ (19.8\%)} & \makecell{123 \\ (25.1\%)} & \makecell{81 \\ (52.9\%)} & \makecell{118 \\ (37.3\%)} 
   & \makecell{3.51 \\ (28.6\%)} & \makecell{8.47 \\ (69.1\%)} & \makecell{3.41 \\ (41.37\%)} & \makecell{8.46 \\ (69.47\%)} \\
 \hline
 3 & \makecell{10 \\ (2.4\%)} & \makecell{6 \\ (1.22\%)} & \makecell{9 \\ (5.9\%)} & \makecell{6 \\ (1.9\%)} 
   & \makecell{1.6 \\ (12.9\%)} & \makecell{1.51 \\ (12.3\%)} & \makecell{1.59 \\ (19.27\%)} & \makecell{1.51 \\ (12.36\%)} \\
 \hline
 4 & \makecell{11 \\ (2.24\%)} & \makecell{11 \\ (2.24\%)} & \makecell{11 \\ (7.2\%)} & \makecell{11 \\ (3.5\%)} 
   & \makecell{0.07 \\ (0.57\%)} & \makecell{0.07 \\ (0.57\%)} & \makecell{0.07 \\ (0.85\%)} & \makecell{0.07 \\ (0.57\%)} \\
 \hline
\end{tabular}
\vspace{.1cm}
\caption{Exclusive transactions by sender, grouped by how many distinct builders each sender interacts with. 
Columns distinguish senders defined by \texttt{from} vs. \texttt{to} addresses, with and without a minimum threshold ($\geq 5$ transactions). The left panel reports the count \textbf{of exclusive transactions that included on chain}, with the share of total exclusive transactions included on-chain underneath. The right panel reports total fees in ETH \textbf{among the transactions included on chain}, with the share of total underneath.}
\label{table:exclusive_relationship_and_fees-2}
\end{table}

Tables~\ref{table:exclusive_relationship_and_fees} and~\ref{table:exclusive_relationship_and_fees-2} group transactions by the number of builders each sender interacts with. We use both definitions of ``sender'' discussed in the previous subsection—\texttt{to} and \texttt{from} addresses—and present results both with and without filtering out senders who submit fewer than five transactions.\footnote{This filter helps address a possible mechanical biases: a sender with only few transactions is more likely to appear to interact with just one builder, which may reflect data sparsity rather than a true exclusive relationship.} Table~\ref{table:exclusive_relationship_and_fees} reports transaction counts and the total fees paid, while Table~\ref{table:exclusive_relationship_and_fees-2} restricts the analysis to transactions that were ultimately included on-chain.

In Table~\ref{table:exclusive_relationship_and_fees}  transaction counts vary somewhat with the definition of ``sender'' but the fee data are more robust: by value, roughly half of all exclusive transactions arise from exclusive relationships between senders and builders. However, once we restrict to on-chain transactions, this share drops sharply. Table \ref{table:exclusive_relationship_and_fees-2} shows that if we use \texttt{to} addresses---which better identify searchers---then only around 7\% of the value of exclusive transactions that land on-chain can be traced to exclusive relationships.  Hence,  exclusive relationships between senders and builders play only a limited role in explaining why, as noted earlier, 85\% of the value of on-chain transactions comes from exclusive transactions.\footnote{
Tables~\ref{table:exclusive_relationship_and_fees_without_exclusvie} and \ref{table:exclusive_relationship_and_fees-2-new-exclusive} in Appendix replicate Tables~\ref{table:exclusive_relationship_and_fees} and~\ref{table:exclusive_relationship_and_fees-2}, but excluding from the set of exclusive transactions those that we previously identified as ``partially exclusive'' or ``similar transactions''. The main result is unchanged: under this alternative categorization, only around 8.4\% of the value of exclusive transactions that land on-chain can be traced to exclusive relationships.}

\begin{table}[thb]
\centering
\begin{tabular}{||l||c|c|c||c|c|c||}
 \hline
 \multirow{2}{*}{Builder} 
   & \multicolumn{3}{c||}{All senders} 
   & \multicolumn{3}{c||}{Senders with $\geq$5 exclusive txs} \\ 
 \cline{2-7}
   & \makecell{Exclusive \\ txs} & \makecell{From  \\ (adds/txs)} & \makecell{To  \\ (adds/txs)} 
   & \makecell{ Exclusive \\ txs} & \makecell{From  \\ (adds/txs)} & \makecell{To  \\ (adds/txs)} \\
 \hline\hline
 Titan Builder & 3106 & 360 / 2519 & 181 / 2457 & 2891 & 20 / 2115 & 15 / 2242 \\
 Rsync Builder & 1335 & 79 / 220 & 20 / 50 & 1305 & 13 / 105 & 1 / 14 \\
 Ty for the Block & 304 & 3 / 304 & 5 / 304 & 300 & 1 / 300 & 3 / 300 \\
 Boba Builder & 148 & 9 / 146 & 7 / 144 & 137 & 2 / 136 & 1 / 136 \\
 Builder0x69 & 69 & 9 / 15 & 5 / 7 & 59 & 1 / 5 & 0 / 0 \\
 Flashbots & 61 & 17 / 19 & 10 / 10 & 48 & 0 / 0 & 0 / 0 \\
 Gambit Labs & 54 & 3 / 51 & 3 / 51 & 52 & 1 / 49 & 1 / 49 \\
 Bob the Builder & 31 & 5 / 11 & 2 / 7 & 26 & 1 / 6 & 1 / 6 \\
 I Can Haz Block? & 31 & 1 / 24 & 0 / 0 & 31 & 1 / 24 & 0 / 0 \\
 Blockbeelder.com & 30 & 0 / 0 & 0 / 0 & 30 & 0 / 0 & 0 / 0 \\
 Geth Imposter & 21 & 3 / 21 & 3 / 21 & 17 & 2 / 17 & 2 / 17 \\
 Beaverbuild.org & 14 & 0 / 0 & 0 / 0 & 13 & 0 / 0 & 0 / 0 \\
 \makecell[l]{Buildernet \\~~~(Nethermind) } & 11 & 2 / 2 & 0 / 0 & 10 & 0 / 0 & 0 / 0 \\
 Payload.de & 7 & 3 / 3 & 0 / 0 & 4 & 0 / 0 & 0 / 0 \\
 Smithbot.xyz & 2 & 2 / 2 & 2 / 2 & -- & -- & -- \\
 Mars & 2 & 0 / 0 & 0 / 0 & 2 & 0 / 0 & 0 / 0 \\
 Bloxroute & 1 & 0 / 0 & 0 / 0 & 1 & 0 / 0 & 0 / 0 \\
 Penguinbuild.org & 1 & 1 / 1 & 0 / 0 & 1 & 0 / 0 & 0 / 0 \\
 \hline
\end{tabular}
\vspace{.1cm}
\caption{Number of exclusive transactions by builder, along with the number of distinct \texttt{from} and \texttt{to} addresses that sent exclusive transactions only to that builder and the number of transactions they sent. The left panel includes all senders; the right panel restricts to senders with at least 5 exclusive transactions.\label{table:exclusive_by_builder}}
\end{table}
 
Table~\ref{table:exclusive_by_builder} performs a similar analysis as Table \ref{table:exclusive_relationship_and_fees}, but builder by builder. It reports the number of exclusive transactions each builder receives and the extent to which these come from senders who interact only with that builder (under different definitions of ``sender''). The key insight is that the importance of exclusive relationships varies, both in absolute numbers and in the percentage of exclusive transactions received: the number of exclusive relationships goes from at least 15 (Titan builder) to zero; the fraction of exclusive transactions from exclusive relationships goes from 100\% (Ty for the Block) to, again, zero.

\subsection{Transactions' transitions.}

We next ask whether transaction status evolves over time. For this purpose, we define each transaction’s status (public, private, or exclusive) \textit{within each bidding cycle} and then track how that status changes across cycles.

Among the 1,330 transactions observed across multiple bidding cycles, public transactions remain public in all subsequent cycles. Only a handful of private transactions (13 out of 208) later appear in the public mempool.  By contrast, the majority of exclusive transactions transition to private or even public. Of the transactions that begin as exclusive, 162 remain exclusive across all observed cycles, while 174 transition to private.\footnote{We also find 33 transactions that were initially exclusive to a builder and later included in a subsequent cycle won by the same builder. Because those cycles are not in our primary dataset, we cannot determine whether the transactions remained exclusive or became private.} In addition, 12 transactions that were initially exclusive eventually surfaced in the public mempool. Nine of these were exclusive for two cycles and three for three cycles. In four cases, after appearing in the public mempool, these transactions were added on chain by the same builder that had them as exclusive to start with (Titan). All appear to be user-generated: three used the 0x router and four the 1inch router. The existence of these transactions is problematic, as they were first delayed and later exposed in the public mempool, leaving users vulnerable to attacks (and one of them appears to have been sandwiched).\footnote{These transactions are:
\begin{itemize}\vspace{-.35cm}
\item \texttt{0x1daea1584d684385fb25209bad3a49c54a1e2c500e27543334c7b3e695a12ffb},
\item \texttt{0x6605c1c36fcd3d696844519d6ff7199fe4ddd34fb4f3e0ed4da71aedfcf1c633},
\item \texttt{0x3dadbab8e85b6e48a2c33f9029ea4ebebdd3def8b9d907faff15ef81527de968},
\item \texttt{0x9c7364eddb94fe659bc8104585ff86a8a234480fe00a9f9aa3eb410e572ebdf2},
\item \texttt{0x9d476dbc524651cbe26bcafab99f6fe40a823a5588fa5cd5002e5b414a8a2f8c},
\item \texttt{0x417440b083bdcb529663c6615c32d89a92beb7863f1a5d6b12bca5b94f4a0f87} (sandwiched according to \url{https://www.zeromev.org/block?num=21322643}),
\item \texttt{0xe56caee27e79dee8a4c7ed6f945ccbc2cc733d761e19bd68329dc364c61e73a7},
\item \texttt{0xae5b88bc0bffa04811eda0dfdcd7224599e2757339e18b2d57f4f40d61902a4c},
\item \texttt{0xd41139b67a193045b4d60266c084f6dadd0cb95dccd7b2a06ea68653d6a313ec},
\item \texttt{0x597fc0cb0610d1e3ea2baf0f6f90d0d741a6429bfdd9b375bfd2493b70d41dcd},
\item \texttt{0x44e8105d5c672195877dc1ed970511e0f0a15fc25cbcafe71d51d428d54c215d},
\item \texttt{0xd74718e6d8d58bc33334c0c2f6642d9dec49496329329d8fdcb1e698e81abc94}.
\end{itemize}
}

A change in transaction status may also occur \textit{within} a bidding cycle. For instance, a builder unlikely to win may share its exclusive transactions with a competitor that has a better chance. To study this, we focus on Rsync and Titan, the two most prominent builders in our dataset. We define a ``new transaction'' as one included in a block submitted at least one second after a builder’s initial block, and not present in that first submission. For each such new transaction from Titan and Rsync, we check whether it appeared in another builder’s block at least one second earlier.  

We identify 20 cases: 3 involving Rsync and 17 involving Titan. Of Rsync’s three, two were initially exclusive to Beaverbuild and later included in a winning Rsync block. The third---a searcher’s swap---appeared in blocks from five different builders, but only entered Rsync’s block four seconds later (Rsync won that auction).

Of Titan’s 17 cases, 8 came through the Flashbots private mempool. Five originated from a single searcher address\footnote{\texttt{0x89a99a0a17d37419f99cf8dc5ffa578f3cdb58b5}} and appear in Flashbots Protect’s dataset as shared with more than 15 builders. These transactions remained exclusive to Flashbots Builder for about 2.6 seconds before entering Titan’s blocks. Another five transactions from the same address follow the same pattern (initial exclusivity to Flashbots, later appearance in Titan’s blocks) but were never included on-chain. It is plausible that they were also shared via Flashbots Protect, which records only transactions that are eventually included.  

In summary, transaction sharing between builders during the bidding cycle does occur but is relatively rare in our data. When it does occur, it involves transactions submitted via Flashbots Protect that are initially exclusive to the Flashbots builder. We speculate that Flashbots may share such transactions with other builders once it becomes apparent that its own builder is unlikely to win the auction.

\section{Conclusion}

Together with its companion paper~\cite{canidio2025becoming}, this study provides the first systematic analysis of non-winning Ethereum blocks, enabling us to distinguish between public, private, and exclusive transactions. We show that although exclusive transactions represent a minority of all transactions included on chain, they account for the vast majority of builders' revenues---between 77.2\% and 84\% of total fees in winning blocks, depending on the methodology. While exclusive relationships between senders and builders are common, they explain only a small share (approximately 7\%) of the total value of exclusive transactions included on-chain. By demonstrating that exclusive transactions are responsible for the vast majority builders' revenues, our results support the view that exclusivity is a key driver of the PBS auction dynamics and the resulting dominance of a few builders.

We also find that transactions initially classified as exclusive may later change status to private or even public. Specifically, some transactions submitted via Flashbots Protect are exclusive to the Flashbots builder during the early part of a bidding cycle but are later shared with other builders, thereby transitioning to private. In addition, we identify 12 transactions that remain exclusive to a single builder for several bidding cycles before eventually appearing in the public mempool. These transactions are delayed and then exposed to potential attacks.

\bibliography{bib}

\begin{thebibliography}{}

\bibitem[\protect\citeauthoryear{Azar, Casillas, and Farboodi}{Azar
  et~al.}{2025}]{azar2024information}
Azar, P.~D., A.~Casillas, and M.~Farboodi (2025).
\newblock Natural centralization in decentralized finance.
\newblock Technical report, Working paper.

\bibitem[\protect\citeauthoryear{Bahrani, Garimidi, and Roughgarden}{Bahrani
  et~al.}{2024}]{bahrani2024centralization}
Bahrani, M., P.~Garimidi, and T.~Roughgarden (2024).
\newblock Centralization in block building and proposer-builder separation.
\newblock {\em arXiv preprint arXiv:2401.12120\/}.

\bibitem[\protect\citeauthoryear{Canidio and Pahari}{Canidio and
  Pahari}{2025}]{canidio2025becoming}
Canidio, A. and V.~Pahari (2025).
\newblock Becoming immutable: How ethereum is made.
\newblock {\em arXiv preprint arXiv:2506.04940\/}.

\bibitem[\protect\citeauthoryear{Capponi, Jia, and Olafsson}{Capponi
  et~al.}{2024}]{capponi2024proposer}
Capponi, A., R.~Jia, and S.~Olafsson (2024).
\newblock Proposer-builder separation, payment for order flows, and
  centralization in blockchain.
\newblock {\em Payment for Order Flows, and Centralization in Blockchain
  (February 12, 2024)\/}.

\bibitem[\protect\citeauthoryear{Capponi, Jia, and Wang}{Capponi
  et~al.}{2025}]{CAPPONI2025104132}
Capponi, A., R.~Jia, and K.~Y. Wang (2025).
\newblock Maximal extractable value and allocative inefficiencies in public
  blockchains.
\newblock {\em Journal of Financial Economics\/}~{\em 172}, 104132.

\bibitem[\protect\citeauthoryear{Gupta, Pai, and Resnick}{Gupta
  et~al.}{2023}]{gupta2023centralizingeffectsprivateorder}
Gupta, T., M.~M. Pai, and M.~Resnick (2023).
\newblock The centralizing effects of private order flow on proposer-builder
  separation.

\bibitem[\protect\citeauthoryear{Janicot and Vinyas}{Janicot and
  Vinyas}{2025}]{janicot2025privatemevprotectionrpcs}
Janicot, P. and A.~Vinyas (2025).
\newblock Private mev protection rpcs: Benchmark stud.

\bibitem[\protect\citeauthoryear{Kilbourn}{Kilbourn}{2022}]{Quintus}
Kilbourn, Q. (2022).
\newblock order flow, auctions and centralisation i - a warning.
\newblock
  \url{https://writings.flashbots.net/order-flow-auctions-and-centralisation}.
\newblock Accessed: 2025-11-04.

\bibitem[\protect\citeauthoryear{Lehar and Parlour}{Lehar and
  Parlour}{2023}]{lehar2023battle}
Lehar, A. and C.~A. Parlour (2023).
\newblock Battle of the bots: Flash loans, miner extractable value and
  efficient settlement.
\newblock {\em Miner Extractable Value and Efficient Settlement (March 8,
  2023)\/}.

\bibitem[\protect\citeauthoryear{{\"O}z, Sui, Thiery, and Matthes}{{\"O}z
  et~al.}{2024}]{oz2024wins}
{\"O}z, B., D.~Sui, T.~Thiery, and F.~Matthes (2024).
\newblock Who wins ethereum block building auctions and why?
\newblock {\em arXiv preprint arXiv:2407.13931\/}.

\bibitem[\protect\citeauthoryear{Thiery}{Thiery}{2023}]{thiery2023empirical}
Thiery, T. (2023).
\newblock Empirical analysis of builders’ behavioral profiles (bbps).
\newblock
  \url{https://ethresear.ch/t/empirical-analysis-of-builders-behavioral-profiles-bbps/16327}.
\newblock Accessed: 2025-11-04.

\bibitem[\protect\citeauthoryear{Titan and {Frontier Research}}{Titan and
  {Frontier Research}}{2023}]{frontier}
Titan and {Frontier Research} (2023).
\newblock {Builder Dominance and Searcher Dependence}.
\newblock
  \url{https://frontier.tech/builder-dominance-and-searcher-dependence}.
\newblock Accessed: 2025-11-04.

\bibitem[\protect\citeauthoryear{Wu, Thiery, Leonardos, and Ventre}{Wu
  et~al.}{2024a}]{wu2024strategic}
Wu, F., T.~Thiery, S.~Leonardos, and C.~Ventre (2024a).
\newblock Strategic bidding wars in on-chain auctions.
\newblock In {\em 2024 IEEE International Conference on Blockchain and
  Cryptocurrency (ICBC)}, pp.\  503--511. IEEE.

\bibitem[\protect\citeauthoryear{Wu, Thiery, Leonardos, and Ventre}{Wu
  et~al.}{2024b}]{wu2024compete}
Wu, F., T.~Thiery, S.~Leonardos, and C.~Ventre (2024b).
\newblock To compete or collude: Bidding incentives in ethereum block building
  auctions.
\newblock In {\em Proceedings of the 5th ACM International Conference on AI in
  Finance}, pp.\  813--821.

\bibitem[\protect\citeauthoryear{Yang, Nayak, and Zhang}{Yang
  et~al.}{2025}]{yang2025decentralization}
Yang, S., K.~Nayak, and F.~Zhang (2025).
\newblock Decentralization of ethereum's builder market.
\newblock In {\em 2025 IEEE Symposium on Security and Privacy (SP)}, pp.\
  1512--1530. IEEE.

\end{thebibliography}
\bibliographystyle{chicago}

\newpage

\appendix

\section{Additional tables}

\begin{table}[h]
\centering
\begin{tabular}{||c||c|c|c|c||c|c|c|c||} 
 \hline
 \multirow{2}{*}{\# of builders} 
   & \multicolumn{4}{c||}{Transactions} 
   & \multicolumn{4}{c||}{Fees (ETH)} \\ 
 \cline{2-9}
   & From & To & From $\geq$5 & To $\geq$5 
   & From & To & From $\geq$5 & To $\geq$5 \\ 
 \hline\hline
1& \makecell{ 3552 \\ (91.22\%)}& \makecell{ 3098 \\ (79.56\%)}& \makecell{ 2923 \\ (82.05\%)}& \makecell{ 2803 \\ (100.0\%)}& \makecell{ 88.7 \\ (82.43\%)}& \makecell{ 68.52 \\ (63.68\%)}& \makecell{ 64.36 \\ (79.56\%)}& \makecell{ 60.59 \\ (61.7\%)} \\ 
\hline
2& \makecell{ 225 \\ (5.78\%)}& \makecell{ 563 \\ (14.46\%)}& \makecell{ 159 \\ (82.05\%)}& \makecell{ 531 \\ (100.0\%)}& \makecell{ 11.32 \\ (10.52\%)}& \makecell{ 29.95 \\ (27.83\%)}& \makecell{ 8.97 \\ (11.09\%)}& \makecell{ 28.46 \\ (28.98\%)} \\ 
\hline
3& \makecell{ 91 \\ (2.34\%)}& \makecell{ 21 \\ (0.54\%)}& \makecell{ 87 \\ (82.05\%)}& \makecell{ 21 \\ (100.0\%)}& \makecell{ 5.5 \\ (5.11\%)}& \makecell{ 3.85 \\ (3.58\%)}& \makecell{ 5.48 \\ (6.77\%)}& \makecell{ 3.85 \\ (3.92\%)} \\ 
\hline
4& \makecell{ 16 \\ (0.41\%)}& \makecell{ 197 \\ (5.06\%)}& \makecell{ 16 \\ (82.05\%)}& \makecell{ 197 \\ (100.0\%)}& \makecell{ 1.26 \\ (1.17\%)}& \makecell{ 4.06 \\ (3.77\%)}& \makecell{ 1.26 \\ (1.56\%)}& \makecell{ 4.06 \\ (4.13\%)} \\ 
\hline
\end{tabular}
\vspace{.1cm}
\caption{Exclusive transactions by sender, grouped by how many distinct builders each sender interacts with. Here \textbf{we exclude from the set of exclusive transactions those we identified as similar or partially exclusive}.
Columns distinguish senders defined by \texttt{from} vs. \texttt{to} addresses, with and without a minimum threshold ($\geq 5$ transactions). The left panel reports transaction counts with the share of total transactions underneath. The right panel reports total fees in ETH with the share of total fees underneath. }
\label{table:exclusive_relationship_and_fees_without_exclusvie}
\end{table}

\begin{table}[h]
\centering
\begin{tabular}{||c||c|c|c|c||c|c|c|c||} 
 \hline
 \multirow{2}{*}{\# of builders} 
   & \multicolumn{4}{c||}{Transactions} 
   & \multicolumn{4}{c||}{Fees (ETH)} \\ 
 \cline{2-9}
   & From & To & From $\geq$5 & To $\geq$5 
   & From & To & From $\geq$5 & To $\geq$5 \\ 
 \hline\hline
1& \makecell{ 384 \\ (93.89\%)}& \makecell{ 334 \\ (81.66\%)}& \makecell{ 66 \\ (23.47\%)}& \makecell{ 165 \\ (76.28\%)}& \makecell{ 8.62 \\ (76.13\%)}& \makecell{ 0.95 \\ (8.39\%)}& \makecell{ 2.01 \\ (41.75\%)}& \makecell{ 0.87 \\ (7.15\%)} \\ 
\hline
2& \makecell{ 19 \\ (4.65\%)}& \makecell{ 66 \\ (16.14\%)}& \makecell{ 20 \\ (23.47\%)}& \makecell{ 125 \\ (76.28\%)}& \makecell{ 0.09 \\ (0.79\%)}& \makecell{ 7.74 \\ (68.36\%)}& \makecell{ 0.13 \\ (2.7\%)}& \makecell{ 8.53 \\ (70.11\%)} \\ 
\hline
3& \makecell{ 4 \\ (0.98\%)}& \makecell{ 3 \\ (0.73\%)}& \makecell{ 5 \\ (23.47\%)}& \makecell{ 3 \\ (76.28\%)}& \makecell{ 1.5 \\ (13.25\%)}& \makecell{ 1.49 \\ (13.16\%)}& \makecell{ 1.54 \\ (31.99\%)}& \makecell{ 1.49 \\ (12.25\%)} \\ 
\hline
4& \makecell{ 2 \\ (0.49\%)}& \makecell{ 6 \\ (1.47\%)}& \makecell{ 5 \\ (23.47\%)}& \makecell{ 19 \\ (76.28\%)}& \makecell{ 1.12 \\ (9.89\%)}& \makecell{ 1.14 \\ (10.07\%)}& \makecell{ 1.13 \\ (23.47\%)}& \makecell{ 1.28 \\ (10.52\%)} \\ 
\hline
\end{tabular}
\vspace{.1cm}
\caption{Exclusive transactions by sender, grouped by how many distinct builders each sender interacts with. Here \textbf{we exclude from the set of exclusive transactions those we identified as similar or partially exclusive}.
Columns distinguish senders defined by \texttt{from} vs. \texttt{to} addresses, with and without a minimum threshold ($\geq 5$ transactions). The left panel reports the count \textbf{of exclusive transactions included on-chain}, with the share of total exclusive transactions included on-chain underneath. The right panel reports total fees in ETH \textbf{among the transactions included on-chain}, with the share of total underneath.}
\label{table:exclusive_relationship_and_fees-2-new-exclusive}
\end{table}

\end{document}